\documentstyle[11pt,aaspp4,flushrt]{article}
 
\newbox\grsign \setbox\grsign=\hbox{$>$} \newdimen\grdimen \grdimen=\ht\grsign
\newbox\simlessbox \newbox\simgreatbox
\setbox\simgreatbox=\hbox{\raise.5ex\hbox{$>$}\llap
     {\lower.5ex\hbox{$\sim$}}}\ht1=\grdimen\dp1=0pt
\setbox\simlessbox=\hbox{\raise.5ex\hbox{$<$}\llap
     {\lower.5ex\hbox{$\sim$}}}\ht2=\grdimen\dp2=0pt

\def\simless{\mathrel{\copy\simlessbox}}

\def\vol#1  {{{#1}{\rm,}\ }}

\def\aj{{AJ}, }  
\def\apj{{ApJ}, } 
\def\apjs{{ApJS}, } 
\def\pasp{{PASP}, }  
\def\mnras{{MNRAS}, } 
\def\aa{{A\&A}, }     
\def\etal{{\it et al.}\ }

\begin{document}
\title{A Study of Nine High-Redshift Clusters of Galaxies : I. The Survey}
\author{J. B. Oke}
\affil{Palomar Observatory, California Institute of Technology, Pasadena, 
CA 91125}
\affil{and}
\affil{Dominion Astrophysical Observatory, 5071 W. Saanich Road, Victoria, 
BC V8X 4M6}
\affil{Electronic mail: oke@dao.nrc.ca}
\author{Marc Postman}
\affil{Space Telescope Science Institute\altaffilmark{1}, 3700 San Martin Drive, 
Baltimore, MD 21218}
\affil{Electronic mail: postman@stsci.edu}
\author{Lori M. Lubin\altaffilmark{2,3}}
\affil{Observatories of the Carnegie Institution of Washington, 813 Santa 
Barbara Street} 
\affil{Pasadena, CA 91101}
\affil{Electronic mail: lml@astro.caltech.edu}

\altaffiltext{1}{Space Telescope Science Institute is operated by the
Association of Universities for Research in Astronomy, Inc.,
under contract to the National Aeronautics and Space Administration.}

\altaffiltext{2}{Hubble Fellow}

\altaffiltext{3}{Present Address : Palomar Observatory, California
Institute of Technology, Mail Stop 105-24, Pasadena, CA 91125}

\vskip 1 cm
\centerline{Accepted for publication in the {\it Astronomical Journal}}

\begin{abstract}

We present a description of the observations and data reduction procedures
for an extensive spectroscopic and multi-band photometric study of
nine high redshift, optically-selected cluster candidates. 
The primary goal of the survey is to establish new constraints 
on cluster and galaxy evolution, with specific emphasis on the 
evolution of galaxy morphology and on the star-formation history
of the galaxies within and around distant clusters.
We have measured 892 new redshifts 
for galaxies with $R \le 23.3$. The data will also serve
as deep probes of the foreground and background large-scale structures.
The observations include broad band optical imaging and
spectroscopy with the Low Resolution Imaging Spectrograph at the 10
meter W. M. Keck Observatory telescope; K-band imaging with 
IRIM at the 4 meter Kitt Peak National Observatory telescope;
and deep, high angular resolution imaging with the WFPC2 onboard the
Hubble Space Telescope. We also describe the procedures used to obtain
morphological information. 
We have established that
six of the nine cluster candidates are indeed
real space density enhancements and are representative
of those typically associated with clusters of galaxies. The remaining
three candidates appear to be projections of
several smaller groups at widely separated distances. This success rate
is consistent with estimates of the false positive rate in 2D optical 
high-$z$ cluster searches.

\end{abstract}


\vfill
\eject

\section{Introduction}

Clusters of galaxies have historically provided an important tool for
studying cosmology and the evolution of galaxies. Because of their
high concentration of galaxies, clusters provide an environment in
which to study large, statistical samples of galaxies. Therefore,
examining clusters of galaxies from the local universe to those at
high redshift allows us to probe galactic evolution to redshifts of
the order of 1. Clusters of galaxies at redshifts of $z \simless 0.2$
have been extremely well cataloged in the optical regime (e.g.\ Abell
1958; Zwicky \etal 1968; Dressler 1980; Shectman 1985; Abell \etal
1989; Lumsden \etal 1992; Dalton \etal 1994). The analyses of local
clusters indicate that these systems are dense (Abell richnesses of
$\sim 30 - 300$ galaxies), massive ($M \sim 10^{14} - 2 \times
10^{15}~h^{-1}~{\rm M_{\odot}}$), and dominated by early-type galaxies
($\sim 50 - 80\%$ of the total galaxy population). These studies
provide a strong basis on which to compare cluster properties at
increasingly higher redshift.

Detailed photometric, spectroscopic and morphological studies have
been extended to clusters of galaxies at redshifts up to $z \sim 0.6$.
The first substantial contribution at these redshifts came from
Butcher \& Oemler (1984) who found a surprisingly large population of
blue galaxies in conjunction with the expected red sequence of
early-type cluster members. Further photometric and spectroscopic
campaigns, including the ambitious CNOC and MORPHS surveys, have
confirmed the progressive bluing of the cluster's galaxy population
and have tracked the passive evolution of the early-type galaxies
(Dressler \& Gunn 1992; Oke, Gunn \& Hoessel 1996; Yee, Ellingson \&
Carlberg 1996; Ellingson \etal 1997; Ellis \etal 1997; Stanford et
al.\ 1995, 1997; and references therein). The Hubble Space Telescope
(HST) has enabled morphological classification of
intermediate-redshift ($z \simless 1$) galaxies on scales which are
comparable to the classifications made of their local
counterparts. These high-resolution studies have revealed that there
may be a substantial change in the morphological composition of the
clusters (Smail \etal 1997; Dressler \etal 1997; however, see Stanford
\etal 1997). All of these results imply that there is a significant
amount of evolution occurring in the cluster environment between
redshifts of $z \approx 0.5$ and $z = 0.0$.  In order to understand this
apparent change in the galaxy population, it is essential to probe in
similar detail clusters of galaxies at even higher redshift where the
effects of evolution and cosmology will be even greater.  In light of
this, we have undertaken an extensive survey of nine candidate
clusters of galaxies at redshifts greater than 0.6. Only a few
optical/near-IR surveys have attempted to detect systematically
clusters at high redshift; therefore, we have chosen our cluster
sample from the Gunn, Hoessel, \& Oke (1984) survey and the Palomar
Distant Cluster Survey (Postman \etal\ 1996).

The observational data compiled for this survey, to date, includes deep $BVRIK$
photometry, over 900 low-resolution spectra, and deep F606W/F702W/F814W
imagery from HST. The large redshift database allows us to reliably
distinguish between physically real clusters and chance line-of-sight
projections. The full data allow us to measure the global properties
of the clusters, such as profile shape and dynamics, as well as the
individual properties of the cluster galaxies, such as color,
star-formation rate, and morphology.  In this introductory paper to
our high-redshift cluster series, we describe in detail the
observational and data reduction techniques of each aspect of this
survey and present the redshift histograms for our nine
fields. The subsequent papers in this series will present the specific
analyses and scientific results of this survey. These papers include
the second and third installments in this series which present a
detailed photometric, spectroscopic, and morphological analyses of the
first two clusters to be completed in this survey, CL0023+0423 and
CL1604+4304 (Postman, Lubin \& Oke 1998; Lubin \etal 1998).

\section{The Cluster Sample}

Over the past several years, clusters of galaxies at intermediate
redshifts ($z \sim 0.5$) have been well--studied (e.g.\ Dressler \&
Gunn 1992; Dressler \etal 1994; Oke, Gunn \& Hoessel 1996; Carlberg
\etal 1997; Ellis \etal 1997; Smail \etal 1997).  With the
completion of the Keck Telescope and the Low Resolution Imaging
Spectrometer (LRIS), the ability to obtain spectra of
cluster galaxies at $z > 0.5$ became feasible with only modest
allocations of observing time.  Two catalogs of distant, northern cluster 
candidates were available, those of Gunn, Hoessel \& Oke (1986) and Postman et
al.\ (1996). Both surveys were based on six fields spread
fairly uniformly around the sky. The present sample consists of
nine candidate high--redshift clusters selected from these fields.  

Seven of the nine clusters are from the catalog of
Gunn, Hoessel \& Oke (1986).  All of these clusters were discovered on
red (600-710 nm) or near-infrared (800 nm) images and had spectra which were 
taken with
either the Prime Focus Universal Extragalactic Instrument (a.k.a. PFUEI)
or the Double Spectrograph on the Hale 5-meter telescope.  The
following six clusters were discovered on the IV-N plates taken with
the Mayall 4-meter telescope. Cluster CL0023+0423 has an unpublished
redshift of 0.832.  Cluster CL0231+0048 has an unpublished redshift of
0.60 or 0.69 and was known to have almost exclusively blue galaxies.
Cluster 0943+4804 has an unpublished redshift of 0.698. Clusters
CL1324+3011, CL1604+4304, and CL1604+4321 have published redshifts of
0.751, 0.895, and 0.918, respectively. Cluster CL2157+0347 was
discovered on red image tube images taken with the Hale 5-meter
telescope and has a published redshift of 0.820 (Gunn, Hoessel \& Oke
1986). Images of some of these clusters can be found in Gunn, Hoessel
\& Oke (1986).

Two clusters, CL1325+3009 and CL1607+4109, were chosen from the
Palomar Distant Cluster Survey (PDCS), a CCD drift-scan survey taken
with the Palomar 5-meter telescope and using the F555W and F785LP
filters (Postman \etal 1996).
The candidate clusters detected in this survey had only estimated
redshifts which were based on the cluster profile and luminosity
function of the galaxies. These two candidates are both
from the supplemental PDCS catalog which means they are of lower
significance but their redshift estimates were close to unity
making them desirable for spectroscopic follow-up.
The estimated redshifts of CL1325+3009 and
CL1607+4109 are 0.9 and 0.8, respectively. 

The final list of clusters is given in Table 1 where the J2000
coordinates are used to identify the clusters (see above). Column 2
lists alternate IDs of the clusters as used by Gunn, Hoessel \& Oke
(1986) and Postman \etal (1996). The J2000 coordinates used for the
cluster observations are given in columns 3 and 4, while the redshifts
of the clusters and other groups in the line-of sight are listed in
column 5 (for the spectroscopic reductions, see Sect.\ 4.2).

\section{The Observations}

\subsection{Keck Broad-Band Observations}

All of the optical observations, both broad band and spectroscopic,
were carried out with the Low Resolution Imaging Spectrometer (LRIS;
Oke \etal 1996) on either the Keck I or Keck II telescopes.  In
imaging mode LRIS has a field of $6 \times 8$ arcminutes.  With the
Tektronix $2048 \times 2048$ back-illuminated and coated CCD with $24
\times 24$ micron pixels, the sampling at the CCD is 4.65 pixels per
arcsecond.  Many of the observations were begun almost immediately
after Keck I and LRIS were commissioned.  Consequently the image
quality was not at the optimum level.  Even so, the full width half
maximum for the images was typically 0.8 to 1.0 arcsecond.

\begin{figure}
\epsscale{1.0}
\plotone{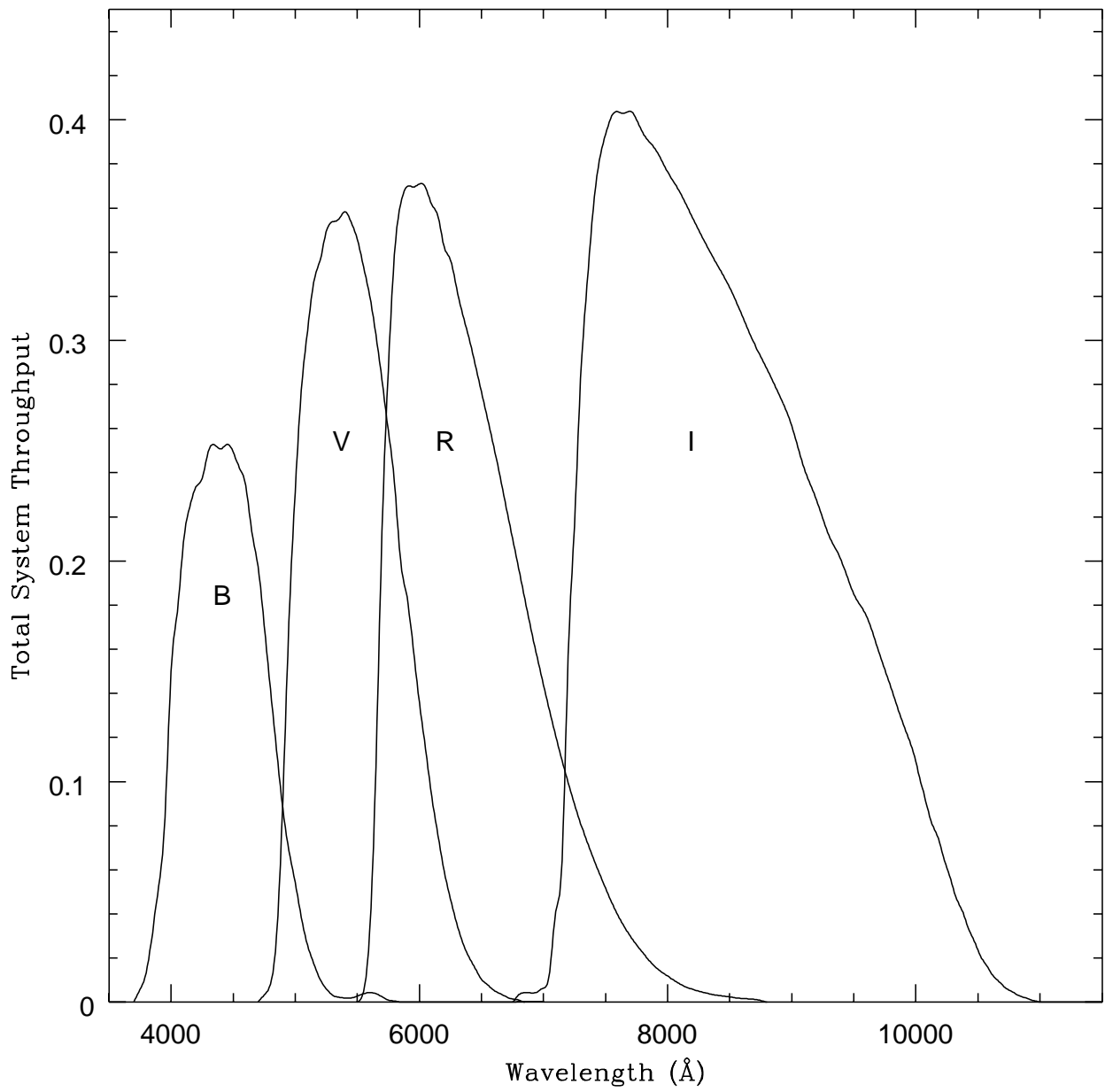}
\caption{The response curves of the optical filter set.}
\label{fig-dqe}
\end{figure}

LRIS has a standard set of BVRI glass filters which match the Cousins
system well.  These filters must go in front of the camera lens and,
consequently, are large.  The B filter, however, is somewhat undersized
because it was impossible to obtain a sufficiently large piece of
Schott BG37 glass.  As a result there is some vignetting in the B
images which must be removed by the flat fielding.  The mean
wavelengths for the four LRIS filters are given in Table 2.  
Figure~\ref{fig-dqe}
shows the total system throughput (optics+detector+filter) for the
LRIS BVRI passbands.

Previous observations at intermediate and high redshift (e.g.\ Butcher
\& Oemler 1984; Rakos \& Schombert 1995) indicate that, at redshifts
approaching unity, the cluster populations may be dominated by blue
galaxies. Consequently, it was desirable to include observations in
the B filter.  Even though the night sky background is low in B, the
total exposure time still had to be long since the quantum efficiency
of the CCD is dropping towards the blue, and faint blue galaxies are
still relatively red objects. At V the quantum efficiency is fairly
high, and the night sky is still low while at R and I the quantum
efficiency is near its peak, but the night sky is becoming bright.

The total exposure times in each filter are listed in Table 2 and were
chosen to give fairly uniform errors in the photometry.  The B, V, and
R observations consisted of two equal exposures to allow for accurate
cosmic ray rejection.  The I exposure time was broken into 3 equal
exposures in order to cope with the cosmic rays, as well as to avoid
approaching too near to the CCD saturation level.  In retrospect it
would have been desirable to obtain about twice the total exposure in
I.  Typical AB magnitudes and fractional standard deviations (${\rm
AB} \pm \sigma_{\rm AB}$) of a representative galaxy at $z = 0.9$ in
each filter are given in columns 4 through 6 of Table 2 .  Column 4
gives these values (as defined below) for a galaxy epoch of 0.7 Gyr,
while the next two columns give similar data for galaxies whose epochs
are 1.4 Gyr and 4.0 Gyr, respectively. These times are for Bruzual and 
Charlot (1993) models with a star formation rate that decreases with 
a time constant of 0.6 Gyr after time zero.  Since errors are dominated by
the sky brightness, the error in each band increases by approximately
a factor of 2.512 when the value of AB increases by 1 magnitude.

The overall program was to obtain both broad-band images for
photometry and slit-mask spectra.  Therefore, if a night was
photometric, the highest priority was given to broad-band photometry.
If the sky was not photometric, spectra were obtained.  
An effort was made to obtain observations in all bands on the same
night.  All observations were made with LRIS rotated to a position
angle on the sky of 180 degrees.  In addition to the exposures, two
kinds of flat fields were obtained.  During each twilight, dome flats
were taken with the telescope pointed to the zenith.  Two nearly
standard overhead projectors mounted on opposite sides of the primary
mirror on the large telescope azimuth ring illuminated a round area on
the surface of the dome directly in front of the primary mirror.  The
projectors have light shields so that only light which goes through
the focusing lens enters the dome.  The dome itself has a rippled
surface on scales of about 100 mm and less and is painted with
aluminum paint.  The result is a fairly uniformly illuminated circle
as seen by the telescope.  The light level is suitable for exposures
through the four BVRI filters.  Flat fields were also obtained by
looking at the twilight sky.  One worry with such flat fields is that
the light is highly polarized and therefore not really representative
of the background night sky seen during the regular exposures.  Since
the flat fielding images were usually not adequate, another approach
described below had to be invoked.

In order to calibrate the observations used to derive BVRI magnitudes
on the standard Cousins-Bessell-Landolt (Cape) system, exposures were
made of Landolt standard star fields (Landolt 1992).  The fields
selected had between 3 and 8 stars spanning a wide range of colors all
located within a single LRIS field of view.  LRIS requires a minimum
exposure time of 2 seconds to maintain uniformity over the field of
view and it was, therefore, necessary to have the telescope out of
focus by 2 mm for the Landolt fields.  The fields observed are
PG0231+051, Rubin 149, PG0918+029, PG1323-086, SA 107, Mark A, and SA
113.  One, and occasionally two, of these fields were observed on each
night that photometric observations were made.

\subsection{Slit-Mask Spectroscopic Observations}

Observations through slit-masks are the standard mode for observing
many faint objects with LRIS.  There is also a long slit option which
is operationally identical to the slit-mask mode and is most useful
for observing individual calibration stars.  All of our spectral
observations were made using the 300 g/mm grating which is blazed at
5000 \AA.  This grating provides a dispersion of 2.38 \AA\ per pixel,
and a spectral coverage of 5100 \AA.  The grating angle was set so
that it provided coverage from approximately 4400 \AA\ to 9500 \AA\ in
the first order.  This range was chosen because it was anticipated
that galaxies would be found with redshifts of at least 1.1, and it
was imperative that the ${\rm O[II]}\lambda3727$ emission line always
be in an accessible part of the spectrum.  A GG495 glass filter was
used to eliminate the overlapping second order spectrum; there is
therefore no second order contamination below 9700 $\AA$.  Because
slits are located at various positions in the focal plane, the
wavelength range covered with any slit will deviate from the nominal
range by up to approximately 600 \AA\ in either direction.

The computer program to design slit-masks (Cohen 1995) can have as an
input either accurate relative values of RA and Dec or positions
measured in pixels from an image taken with LRIS.  The latter input
was used in almost every case.  Additional parameters which are
necessary are (a) position angle on the sky of the slit (always 180
degrees), (b) the appropriate hour angle anticipated, (c) the
individual minimum slit length (8 arcseconds used to provide enough slit
for sky subtraction), and (d) the slit width.  Initially only two
possible slit widths were available, 0.7 and 1.4 arcseconds.  Because of
the limited experience initially available with LRIS, it was decided
to use the 1.4 arcsecond slit to minimize any problem in setting objects
in the slits.  A more ideal 1.0 arcsecond slit is now available and is 
being used for this program.
The optical resolution of the spectrograph is a little less than 2 pixels. 
The 1.0 and 1.4 arcsecond slits, however, correspond to 4.6 and 6.5 
pixels, respectively, so that the spectra are oversampled.  
All spectra were therefore smoothed with a running 
$\left(\matrix{1&2&3&4&3&2&1}\right)$ weighting function.

One of the very early tasks was to obtain images of the clusters for
the purpose of designing the slit-masks.  The R filter was used for
these preliminary cluster images because of the high CCD throughput in
that passband and to optimize detection of distant cluster
galaxies. Exposures of 600 sec in duration were taken under
photometric or near photometric sky conditions.  These were flattened
using the dome flats already described and run through FOCAS (Valdes
1982) programs to generate object magnitudes and positions.  Care was
taken to fix any positions of close pairs not correctly deblended.  An
approximate extinction correction was applied, and the magnitudes
corrected to proper R magnitudes.  The range in right ascension was
restricted to 600 pixels centered on the cluster.  The range in
declination was the full picture height of about 2000 pixels.  A list
of all objects brighter than approximately R = $23.5 \pm 0.1$ in this
box was then extracted in order to provide the list of candidates for
slit-mask spectra. The uncertainty in the magnitude limit is dominated
by flat-field residuals in these early data. The subsequent images
obtained for our photometric analyses flattened to 1\% or
better. Typically, 200 galaxies satisfied the above criteria.  Each
mask can have about 35 slits and, thus, 6 masks were needed for each
cluster. However, the non-uniform galaxy distribution and the desire
to observe some of the fainter targets more than once (for enhanced
signal-to-noise and redshift consistency checks) impose a reduction in
the number of targets observed down to approximately 130 -- 150
objects per cluster field. In early slit-mask setups, for example CL0023+0423
and CL1604+4304, about 30 percent of the objects were 
observed more than once. In later setups about 10 percent of the objects were 
observed more than once.  Two slit locations were always reserved for
setup stars.

The high throughput of LRIS and the excellent image quality meant that
we were able to achieve a signal-to-noise level sufficient to measure
redshifts and study spectral features in a single 3600 sec exposure
for each slit-mask.  In the event of significant cloud coverage, two
such exposures per mask were obtained.

To minimize the effects of instrument flexure, a combined Neon and
Argon discharge lamp exposure was taken immediately following the
object exposure.  In addition, a flat field exposure was made using
two quartz halogen lamp sources.  Both the discharge and quartz
halogen lamp radiation is reflected and scattered off the diffuse
white underside of the dust cover which, when closed, is about 300 mm
in front of the spectrograph slit-mask.  The flat fields obtained in
this way are excellent for flattening spectra and minimize any problems 
caused by flexure.

Since spectra were obtained at small zenith distances and the slits
are fairly broad, it makes sense to try to calibrate the spectra
approximately to relative absolute spectral energy distributions.
This will at least remove the instrumental sensitivity.  Spectra were,
therefore, taken of standard stars using the same instrumental setup
except that a long slit with a width of 1.4 arcseconds was used
instead of a slit-mask.  The standard stars were chosen from the list
of faint HST standards by Oke (1992).  Such spectra were taken at
least once each night.  They were always accompanied by discharge lamp
and quartz halogen lamp exposures.

\subsection{Infrared Observations at KPNO}

For our high--redshift cluster sample, infrared imagery extends our
ability to study the rest-frame optical spectral region which is
dominated by the long-lived, near-solar mass stars, where as the
optical bands sample the poorly observed rest-frame ultraviolet, which
is dominated by the short-lived, massive stars. Specifically, infrared
imagery is well suited to examining the red, ``passive'' early-type
population in high--redshift clusters (Gunn 1990; Arag\'on-Salamanca
\etal 1993; Rakos \& Schombert 1995; Stanford, Eisenhardt \&
Dickinson 1994; Lubin 1996). In addition, K--corrections in the
infrared are smooth, better defined, and almost independent of Hubble
type.
 
Therefore, we are in the process of carrying out deep infrared imaging
of the cluster sample with the Infrared Imaging Camera (IRIM) on the
Mayall 4m telescope at Kitt Peak National Observatory.  IRIM contains
a $256 \times 256$ NICMOS3 HgCdTe array which has a resolution of 0.60
arcsecond~per~pixel on the 4m telescope. The field-of-view ($154
\times 154$ arcseconds) covers the central region of each cluster and
corresponds almost exactly to the field-of-view of our Hubble Space
Telescope (HST) imaging (see Sect.\ 3.4). We have chosen to make the
deepest possible observations over this region and, thus, we have not
mosaiced the entire LRIS field of view.
 
We observe in the $2.2\mu\ K^{'}$ band. This intermediate band filter
avoids the excessive thermal background in the wider $K$
passband. Each central field is observed using a $4 \times 4$ dither
pattern with a stepsize of $10^{''}$ and a total extent of
$30^{''}$. Each exposure has an effective exposure time of 1 minute,
with 4 co-additions of individual, background--limited 15 second
integrations. The total integration time on an individual cluster
varies between 3 and 4.4 hours. Because the fields are not excessively
crowded, we are able to use in-field dithering to create a global sky
flat. Several HST standard stars from the list of Persson \etal (1997)
are observed each night in five separate array positions for each
star. Currently, observations have been made over 3 days in October
1996 and 2.5 days in March 1997.
 
\subsection{HST WFPC2 Observations}

Each of the nine cluster candidates in our
sample has been or will be observed with the Wide Field Planetary
Camera 2 (WFPC2) on the HST.
The HST imaging covers one WFPC2 field of view (roughly $2^{'} \times
2^{'}$) of the central region of each cluster. The specifications of
these observations, including the filters, exposure times, and current
status, are given in Table 3. Two of the clusters, CL1324+3011 and
CL1604+4321, were originally observed by J. Westphal in Cycle 4
(GO-5234) and have been obtained from the HST archive. The
observations of the remaining seven clusters are being conducted in
cycles 5 and 6 of the HST Guest Observer program (GO-6000, 6581; PI
Postman).
 
The HST observations are being used to provide a detailed description
of the structural characteristics of the galaxies detected in the
central region of each cluster.  The HST images for this survey are
processed through the Medium Deep Survey (MDS; Griffiths \etal\ 1994)
data reduction pipeline in order to objectively detect extended
sources and provide a quantitative analysis of their structural
properties. The brightest subsample of these galaxies are visually
classified according to the Revised Hubble system of nearby galaxies
given in e.g.\ the Hubble Atlas (Sandage 1961) and the Carnegie Atlas
of Galaxies (Sandage \& Bedke 1994). This subset corresponds to
approximately 200 galaxies per field 
(corresponding to a limiting magnitude of F702W $\approx 24.7$). 
Two clusters, CL0023+0423 and
CL1604+4304, currently have a completed set of spectroscopy, optical
photometry, and WFPC2 observations. The results of the morphological
studies for these clusters are presented in the third paper of this
series (Lubin \etal 1997, hereafter Paper III).  In this paper, we
present a description of the calibration and reduction of the WFPC2
images, the object detection procedure, and the automated and visual
galaxy classification (see Sect.\ 4.4).

\section{Data Reduction}

\subsection{Keck Broadband Image Reduction}

The LRIS imaging data were reduced in a fairly standard fashion.  All
frames were bias-subtracted and pixel-to-pixel sensitivity variations
were removed using dome flats. Large-scale gradients were removed by
dividing each frame by a normalized two-dimensional spline fit to the
sky values in a sky-flat.  The sky-flat was created by generating a
median image from a stack of frames for each night and passband.  For
the I-band data, a two-dimensional fringe map was also created from
the median filtered image by removing large-scale gradients within the
median image. Fringing was then removed from each I-band frame by
subtracting a suitably scaled version of the fringe map.

Excellent flat-fielding was achievable with the B and V band data with
typical sky variations at the 0.2\% level or less on scales of 20
arcseconds and larger.  The R and I band images obtained prior to 1996
had some residual structure in the sky which may have been caused by
unwanted illumination because the Cassegrain sky shield for Keck I had
not yet been installed. However, the amplitude of these features was
typically at the 1.0\% level (or less) on scales of 20 to 50
arcseconds. The left and right edges of all images were unusable due
to vignetting from optics internal to LRIS. All image photometry was
thus confined to the central 1500 $\times$ 2000 pixel region.

Image registration for a given cluster field was performed by
identifying approximately 10 unsaturated stars (detectable in all 4
passbands) to be used as astrometric reference points. The mean X and
Y offsets of these stars in every frame taken of the cluster were
computed relative to their locations in a fiducial B-band image. All
image data for the cluster were then shifted to match the B band
coordinate system using a flux conserving Lagrangian interpolation
scheme to achieve registration at the sub-pixel level.  Once all
frames of a given cluster were registered to a common coordinate
system, the independent exposures in each passband were co-added to
produce the final four BVRI images.

Absolute astrometry was performed by identifying objects on the LRIS
CCD images that are also detected in digitized versions of the new
Palomar Observatory Sky Survey. The number of such objects which are
both unsaturated on the CCD and bright enough to yield accurate
centroids on the digitized sky survey is small - usually no more than
12 objects.  We then use the plate solution from the POSS-II data to
astrometrically calibrate the CCD exposures. The typical rms error in
our equatorial coordinates is about 0.8 arcsecond.

\subsubsection{Object Photometry and Classification}

We use the FOCAS package to detect, classify, and obtain aperture and
isophotal magnitudes for objects in the co-added BVRI images.  We
generate independent object catalogs from the BVR images. However, we
use the R-band isophotes to determine object parameters in the I-band
images because the sky brightness increases significantly redwards of
7500 \AA\ and object detection would have been severely compromised if
done directly on the I-band image.  Our isophotal detection threshold
is set to be $3\sigma_{\rm sky}$ which corresponds, on average, to
$\mu_B = 27.4$, $\mu_V = 26.7$, and $\mu_R = 25.9$ ${\rm
mag~arcsec^{-2}}$, respectively. A minimum object area constraint of
0.55 ${\rm arcsec^{2}}$ is also imposed for detection.  Aperture
magnitudes are computed using a circular aperture with a radius of 3
arcseconds (14 pixels). The limiting magnitudes are B = 25.1, V =
24.1, R = 23.5, and I = 21.7 for a 5-$\sigma$ detection in our
standard 3 arcsecond radius aperture. An aperture radius of 3 arcseconds
corresponds to 12.8 and 14.9$h^{-1}$ kpc at $z = 0.6$ and $z = 0.9$, 
respectively. This aperture choice is consistent with that used for
faint galaxy photometry in other intermediate redshift cluster analyses
({\it e.g.}, Arag\'on-Salamanca \etal 1993, 
Smail, Ellis \& Fitchett 1995, Barger \etal 1996). 

To get optimal star/galaxy classification accuracy, the PSF is
generated from a manually selected sample of unsaturated stars. We
have found that on very deep frames, the automatic PSF star selection
used in FOCAS can occasionally generate overly broad PSFs -- the
probable result of compact galaxies being included in the
computations. Manual selection avoids this problem. The typical FWHM
seeing in our LRIS image data is 0.96 arcsecond.  About 25\% of the
data had FWHM seeing of 0.8 arcsecond.

Lastly, the objects in the BVRI catalogs are matched with one another
using the {\it match} algorithm in FOCAS. This produces a catalog from
which colors can be generated.
 
\subsubsection{Conversion of BVRI to Absolute Flux AB}

With 4 magnitudes in B, V, R, and I, the easiest and most instructive
way to make comparisons with evolutionary models (e.g.\ Bruzual \&
Charlot 1993) is to convert them to absolute fluxes (AB) where AB is
defined as

\begin{equation}
	AB = -2.5~{\rm log}~f_{\nu} - 48.60 
\end{equation}
and $f_{\nu}$ is the flux in $\rm{erg~cm^{-2}~s^{-1}~Hz^{-1}}$.  One
can define ABB, ABV, ABR, and ABI for each band.

The conversion from BVRI to AB would be easy if one knew the
transmission functions of the four bands, the CCD quantum efficiency,
and the throughput of the telescope and spectrograph.  Then, with a
series of absolute spectral energy distributions for standard BVRI
standard stars of different colors, one could perform integrations
over each band and derive the conversions from BVRI to AB directly.
One could also generate AB values for the evolutionary models which
have constructed absolute energy distributions.  Absolute energy
distributions do exist for a number of BVRI standard stars; however,
no adequate information about the details of the BVRI system exist, so
one cannot easily convert the model magnitudes to the corresponding AB
or BVRI magnitudes.

Therefore, a slightly different technique was used.  The known filter,
atmosphere, telescope transmission, and quantum efficiency functions
for LRIS were used to generate a ``natural'' system which is close to
but not identical with the BVRI system. The energy distributions for
the standard BVRI stars were then used with these functions in order
to generate natural magnitudes called ``bvri.''  Nearly all the
standard BVRI stars used were taken from the list of faint
spectrophotometric standard stars whose energy distributions had been
measured for use with the Hubble Space Telescope (Oke 1992).  BVRI
magnitudes were kindly supplied by Landolt (1996).  The natural bvri
could also be easily generated for any of the evolutionary models
being used.  One could then derive relationships between bvri, BVRI
and the ABs.  The mean wavelengths for the ABs are those listed in
Table 2.  The derived conversions are as follows:

\begin{eqnarray}
	ABV & = & V + 0.01 \\
	ABB & = & B - 0.091(B-V) - 0.09 \\
 	ABR & = & R + 0.034(V-R) + 0.17 \\
	ABI & = & I - 0.121(V-I) + 0.44
\end{eqnarray}

The color terms allow for the fact that the LRIS filters are not
exactly at the standard BVRI wavelengths.  The same transformations
were used to convert absolute spectral energy distributions generated
by evolutionary models to BVRI and ABs.

\subsection{LRIS Spectrophotometric Reductions}

The slit-mask images produced by LRIS consist of about 35 spectra,
stacked vertically. Each spectrum spans the full width of the CCD.
The raw spectroscopic images were bias-subtracted, flat-fielded,
edited to remove cosmic ray hits, and rectified in both the spatial
and dispersion directions.  The uncalibrated spectra were then
extracted after subtracting the sky as measured on both sides of each
object spectrum.  Comparison discharge-lamp spectra were extracted
using the identical procedure and the emission line positions fitted
with a third order polynomial.  The least-squares fit is accurate to
about $0.2 -0.5~\AA$ at any point.  The standard star spectra were
processed identically.  All object spectra were corrected for
atmospheric extinction with the $O_{2}$ and water vapor absorption
handled with a separate extinction law.  The unknown spectra were
converted as well as possible into absolute fluxes AB by dividing the
unknown by the standard star observed flux.

The final spectra are approximately relative absolute energy
distributions. They are approximate and relative because the spectra
are not slitless and individual objects may not be centered precisely
on the slits. Spectra are also sometimes taken in non-photometric
conditions.

The majority of objects were faint and photon shot-noise was
nearly always dominant.  Sky subtraction of
the $\lambda 5577$ or $\lambda\lambda 6300,6363$ feature was always
relatively poor because the lines are extremely strong.  In the near
infrared there are many ${\rm OH^{+}}$ emission bands which are
strong, and sky subtraction is relatively poor.  A further serious
problem above 8000\AA\ is the fringing in the CCD which proved 
difficult to eliminate completely. 
The ${\rm OH^{+}}$ emission bands, the fringing, and the
rapidly decreasing quantum efficiency of the CCD mean that above 8000
\AA\ the spectra rapidly deteriorate in quality.  Only very strong
emission lines can be identified unambiguously.

\subsubsection{Redshift Determination}

Since most of the objects observed are fainter than R=21, their
spectra can be quite noisy; in some cases the continuum has a signal-to-noise
ratio approaching unity as R approaches 23.3.
Emission lines, when present, have a higher
S/N and are relatively easy to detect. Thus, a redshift 
measurement procedure which relies, in
part, on visual inspection was used.  The sky-subtracted
two-dimensional spectrum was first displayed, and the pixel locations
of all strongly suspected emission and absorption lines were recorded.
The criterion for a line to be noted was that it have the proper shape
along the dispersion and also stretch from one side of the spectrum to
the other.  Apparent emission or absorption features near the
positions of strong sky emission lines were left out unless they were
at least partially resolved from the night sky line.  One could then
look at the pattern of emission and absorption lines and infer the
identifications.  The emission lines usually seen were ${\rm
[OII]}\lambda3727$, $\rm H\beta$, ${\rm [OIII]\lambda\lambda 4959,
5007}$, and H$\alpha$, or a subset of these depending on the
redshift. Absorption lines are much harder to see visually and only H
and K were normally seen.  Once emission lines and/or H and K were
identified one could often identify the other strong absorption
features $\lambda$3835 blend, H8, H, K, H$\delta$, the G-band, and
H$\gamma$ in the plotted spectrum. The spectrum was then plotted on a
very expanded scale around each suspected line and the center of each
well defined feature determined.  This could easily be done to a
precision of 1\AA (typical instrumental resolution is $\sim11$\AA).  
A redshift was calculated for each line, and an
average taken of all the detected lines.  In a few objects other
emission lines such as the $\rm [Ne V]\lambda\lambda$ 3869, 3968, $[N
II]\lambda\lambda 6548, 6583$, and $\rm [SII]\lambda\lambda 6716,
6730$ were seen.
 
At redshifts near unity one often runs into the problem that only one
emission line is seen.  The line could be H$\alpha$, $[OIII] \lambda
5007$, $\rm H\beta$, or [OII].  In some cases an identification with
H$\alpha $ can be ruled out because the $\rm H\beta$ / [OIII] group is
not seen in the part of the spectrum where it could easily be
detected.  Similarly, one can rule out $\rm H\beta$ /[OIII] because
[OII] is missing. It was sometimes possible to confirm the
identification with [OII] by the presence of the absorption features
$\rm MgII\lambda\lambda$2795 and 2802 (occasionally resolved) and $\rm
FeII\lambda\lambda 2586, 2600$.
 
\begin{figure}
\epsscale{1.0}
\plotone{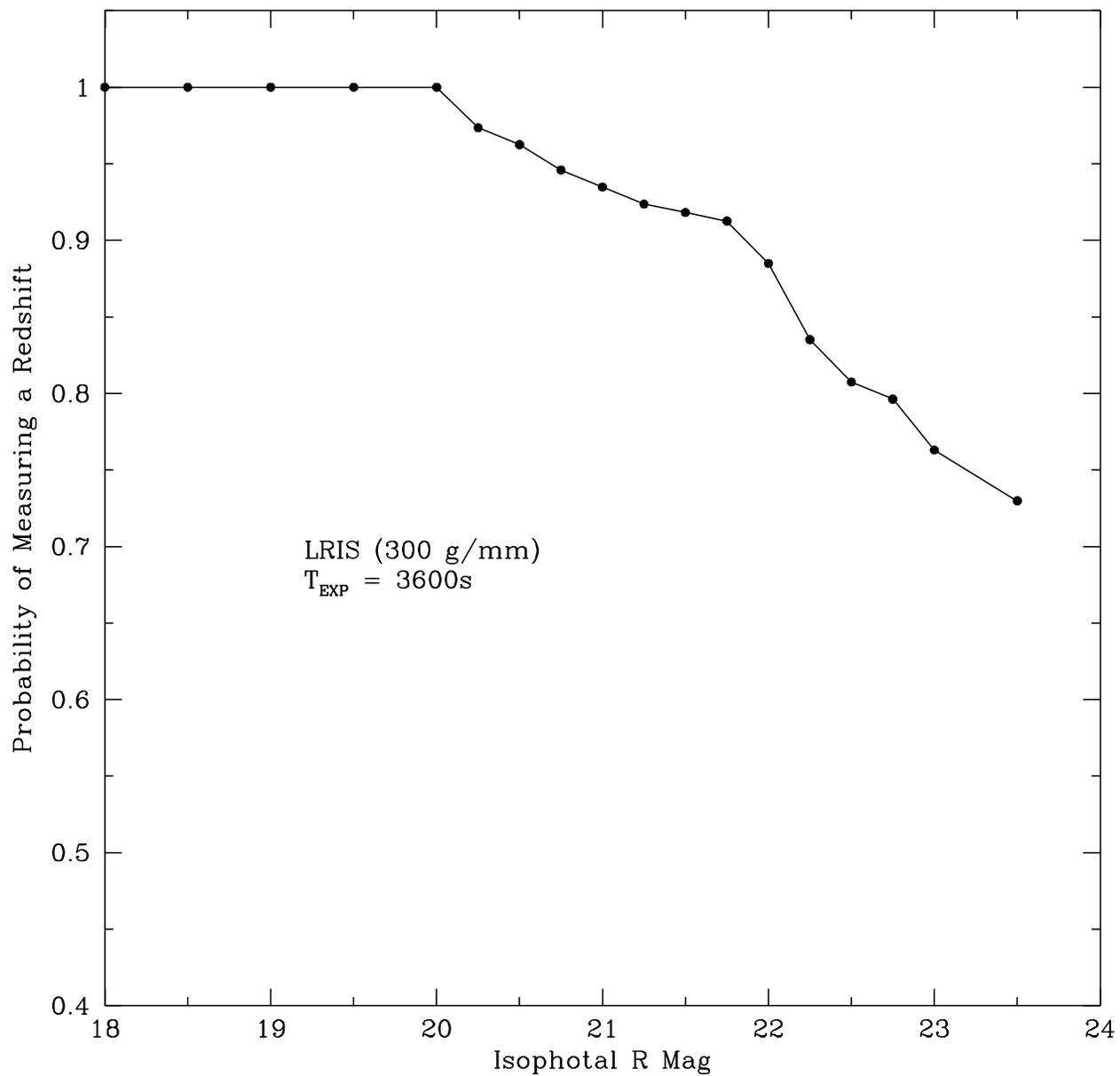}
\caption{Fraction of spectra for which redshifts were successfully
measured as a function of isophotal $R$ magnitude.}
\label{fig-zcomp}
\end{figure}

There are a few spectra where no spectral features are seen even
though the signal-to-noise level is relatively high.  These are
probably cases where $z < 0.2$ or $z > 1.5$ and, thus, the portion of
the spectrum sampled has only very faint features.  In a few cases
where $z < 0.2$ there is evidence for the Na D lines in absorption.
There are also spectra with very low signal-to-noise levels.  In these
cases one can only assume that very strong emission lines are not
present.  The sample of spectra always contained a few faint stars.
Their zero redshifts were usually verified by the presence of the D
lines and H$\alpha$.  Some of the stars were M stars with TiO bands
which were easily identified even in noisy spectra.
Figure~\ref{fig-zcomp} shows our final redshift measurement efficiency
as a function of R magnitude. The probability of measuring a redshift,
either galaxy, quasar, or star, is about 90\% of the objects observed.

If an object is observed more than once, then its
redshift is quite accurate.  In the few cases where multiple
observations of the same galaxy yielded different redshifts, we were
always able to resolve the discrepancy.  In some cases two emission
lines would be seen in one spectrum and only one in a second poorer
spectrum.  In the spectrum with only one line the line was
misidentified.  Occasionally, an object was misidentified.  For the
objects with only a single observation (70 to 90\% depending on the
specific cluster), there is about a 10\% chance
that a gross error has been made in the $z$.  The typical instrumental
error is 0.0006 in redshift which corresponds to
$180~\rm{km~sec^{-1}}$ at $z = 0.0$ and $95~\rm{km~sec^{-1}}$ at $z =
0.90$.

\begin{figure}
\epsscale{1.0}
\plotone{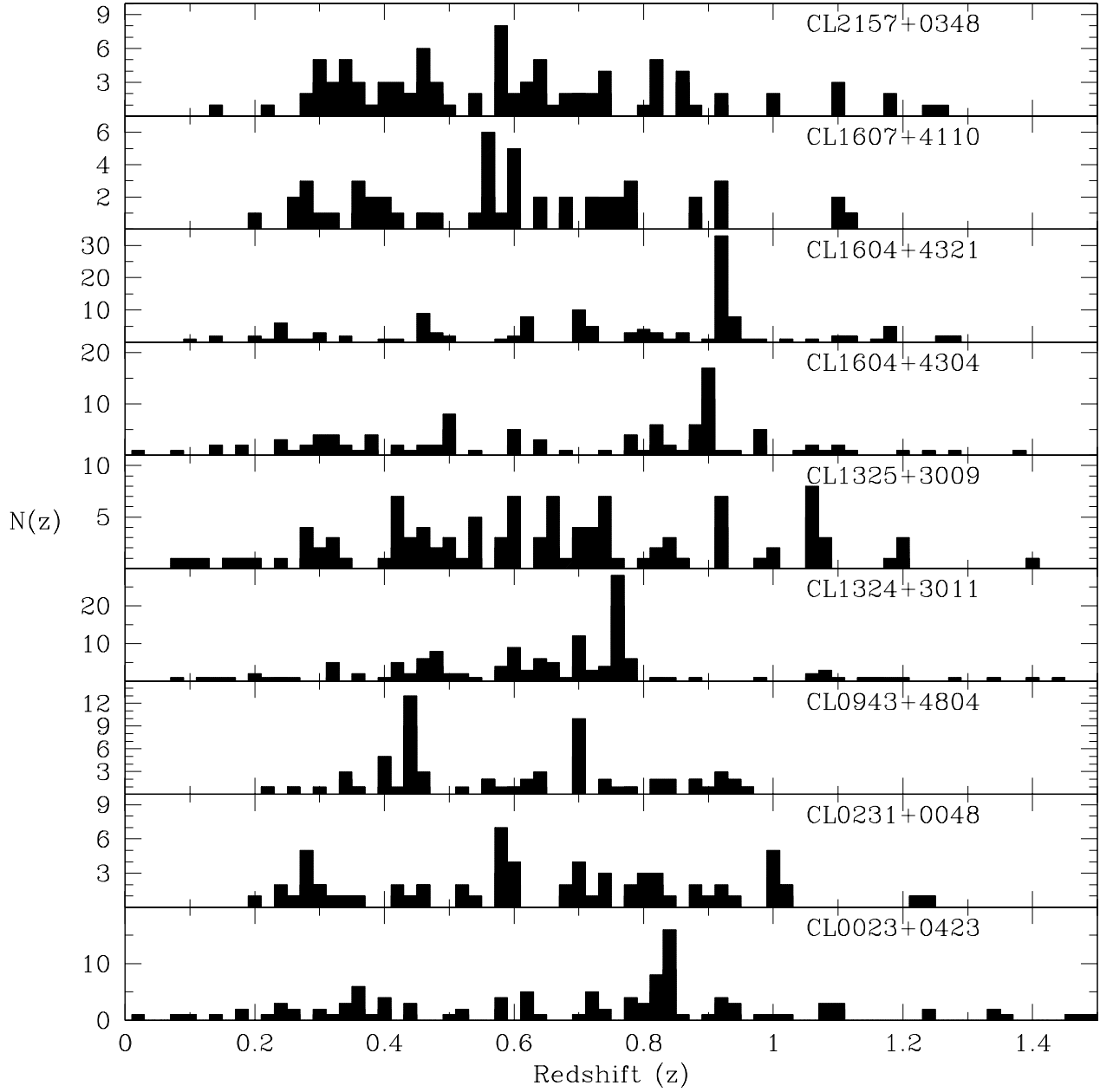}
\caption{The redshift histograms for the nine clusters in
this survey.}
\label{fig-zhist}
\end{figure}

Figure~\ref{fig-zhist} shows the redshift histograms for the
nine clusters in the survey. The redshifts presented here are geocentric
values.  Six of the nine candidate clusters of galaxies are real space
density enhancements.  They are CL0023+0423, CL0943+4804, CL1324+3011,
CL1325+3009, CL1604+4304, and CL1604+4321. The velocity dispersions of
these 6 systems range from 400 km sec$^{-1}$ to 1300 km sec$^{-1}$ and
are typical of clusters at low and intermediate redshifts.  
CL1604+4304 and CL1604+4321 have similar redshift distributions
suggesting the existence of a supercluster system.
A more detailed analysis of the photometric and spectral data of the first
two clusters with complete data, CL0023+0423 and CL1604+4304, are
presented in the second paper of the series (Postman, Lubin \& Oke
1998, hereafter Paper II).

\subsection{IRIM Reduction and Object Detection}

The $K^{'}$ cluster data are reduced using the Deep Infrared Mosaicing
Software (DIMSUM), a publicly available package of IRAF scripts. The
data are linearized, trimmed to exclude masked columns and rows on the
edges of the arrays, and dark--subtracted using dark frames of the same 
exposure length as the observations.  Testing by the authors and
others (M.\ Dickinson and A.\ Stanford) have shown that the best
flat-fielding in $K^{'}$ is obtained with dome flats taken with the
lights off. All images of a given night are flattened by a super flat
made from a series of dome flats taken during the previous day. As
part of the DIMSUM procedure, sky subtraction is done by subtracting a
scaled median of nine temporally adjacent exposures for each frame. 
We have chosen nine exposures for
this procedure as it provides a good statistical sample of the sky
values over a time period where we typically expect the sky variations
to be stable.
A first-pass reduction is used to create an object mask for each frame.
This mask is created from a fully stacked mosaic image. It, therefore,
excludes not only the bright objects, but also those objects too faint
to be detected in an individual exposure. In the second pass, the
object mask is used to avoid object contamination of the sky flat in
the production of sky frames. Final mosaicing of the images of each
cluster are made with a replication of each pixel by a factor of 4 in
both dimensions. This procedure conserves flux while eliminating the
need for interpolation when the individual frames are co-aligned. That
is, there is sufficient resolution that only integer shifts are
necessary when co-adding the individual frames. A bad pixel mask is
used to exclude bad pixels from the final summed images. Object
detection is carried out on the final sum of the $K^{'}$ images.

Object detection, cataloging, and photometry are done using the the
SExtractor image analysis package (Bertin \& Arnouts 1996).  The
isophotal detection threshold is set at the $1.5 \sigma_{\rm sky}$
level which corresponds to, on average, $\mu_{K^{'}} = 22.2~{\rm
mag~arcsec^{-2}}$. A $7 \times 7$ pixel top-hat spatial filter and a
minimum detection area of 35 pixels are also used. Here, ``pixel''
refers to the ``subpixels'' of the original image, i.e. the
$0\farcs{15}$ pixels that result after the $4 \times 4$ replication
described above.  A typical FWHM of a star in the summed image is 1.0
arcsecond. One of us, MP, has explored potential systematic differences
between FOCAS (used for the optical image analysis) and SExtractor. No
significant differences in photometry, astrometry, or classification were 
found providing the object detection parameters are set identically. 
 
Absolute photometric transformations are derived from the observations
of Persson \etal (1997) standard stars. Each standard is observed
five times; first in the center of the array, and then near each of
the four corners. These images are dark-subtracted and flat-fielded in
the same manner described above for the cluster images.  The five
standard star images are then combined into a ``local sky'' frame
using the median percentile clipping option in the IRAF task
IMCOMBINE. The resulting sky image is subtracted from each of the five
standard star images. The instrumental magnitudes of the star are then
measured. The mean instrumental magnitude of the five individual
measurements is used as the assigned instrumental magnitude of that
standard star. The typical standard deviation of these mean
instrumental magnitudes is $0.03$ mag or less. The typical variations
about the nightly photometric transformations are $0.02$ mag or less.
Although a fraction of the observations were carried out in
non-photometric conditions, enough photometric data were obtained for
each cluster to ensure absolute calibration to the required precision
of $< 0.03$ mag.  The conversion from this Vega-based to the AB
magnitude system (see Sect.\ 4.1.2) is given by

\begin{equation}
	ABK^{'}= K^{'} + 1.86
\end{equation}

For this project, we would like to reach a limiting magnitude of
$K^{'} = 20$ for a $5 \sigma$ detection in our standard aperture of
radius $3\farcs{0}$. The limiting magnitudes of our optical
observations (see Sect.\ 4.1.1) indicate that this limit is
appropriate for the bluest cluster members at these
redshifts. Currently, three out of the nine clusters have been
completed down to this limiting magnitude. Future observations to
complete the survey are planned. The first results from the infrared
survey will be presented in Lubin, Oke \& Postman 1998.
 
\subsection{WFPC2 Data Reduction}
 
The Medium Deep Survey (MDS) team has devoted considerable effort to
improving the calibration procedure of WFPC2 images.  Such
improvements are essential to the quantitative analysis of the
faintest extended sources in these images. Since we would like to
extend our analyses to low signal-to-noise ratios, we have used the
final, calibrated images from the MDS reduction pipeline for both the
automated and the visual galaxy classifications of our HST fields. We
briefly describe their data reduction procedure below; however, the
complete details can be found in Ratnatunga, Ostrander \& Griffiths
(1997; hereafter referred to as ROG), Ratnatunga \etal (1994, 1995),
and at the MDS website address
http://astro.phys.cmu.edu/mle/index.html.
 
The WFPC2 images are calibrated using the best available calibration
data.  Presently, these data come from the calibration files created
by the Space Telescope Science Institute (STScI) to calibrate the
Hubble Deep Field (HDF; Williams \etal 1996). Therefore, the static
mask, super-bias, super-dark, and flat field calibration files of the
HDF are adopted for this procedure. Hot pixel tables from STScI for
the given observation period are used to correct warm pixels. Hot
pixels which cannot be corrected to an acceptable accuracy, saturated
pixels, and pixels with a large dark current are ignored. The MDS
software is designed to handle excluded pixels.
 
A corrected version of the standard IRAF/STSDAS combine task is used
to create a stacked image (see ROG and Ratnatunga \etal 1994 for the
specifics of this stacking procedure and the associated statistical
errors). The individually calibrated images are stacked with shifts
that correspond to the nearest integer number of pixels. Slight
variations in the mean sky background are taken into account by
allowing for a mode offset between pointings. This procedure produces
the final reduced image used for the morphological studies. We have
verified the MDS procedure by reducing the HST observations separately
using the standard STSDAS routines. The relative positions and fluxes of
over 100 galaxies per field were compared. No significant differences
were observed; the scatter was consistent within the statistical
uncertainties.
 
\subsubsection{Object Detection in WFPC2 Images}
 
After the images are stacked, the automated object detection is then
performed using a ``find'' algorithm developed by the MDS team
specifically for WFPC data. Because the algorithm does not perform any
pre-convolution of the data, it is insensitive to hot or missing pixel
values (see above). The detection algorithm is based on finding local
maxima in the field. Nearby pixels are then mapped and associated with
that particular maxima. Detections significantly above the noise are
cataloged, and a ``mask'' image is made. A canonical radius of
$0.5^{''}$ is chosen as the object resolution, so that individual
galaxies are not split into smaller components.  All maxima within
this radius are merged with the brightest center.
 
The resulting mask is examined for bright objects which have been
over-resolved and for ghost images or bright stellar diffraction spikes
which have been mistakenly detected as objects. These spurious
detections are interactively flagged for rejection or merger with the
central image. This interactive procedure follows a well-defined set
of guidelines which have been adapted for WFPC2 data. For more details
on this procedure, see the references given above.
 
\subsubsection{Automated Object Classification in WFPC2 Images}
 
The last part of the MDS pipeline is a two-dimensional maximum
likelihood image analysis procedure which automatically optimizes a
model and the number of parameters to be fit to each object image. 
Two scale-free, axisymmetric models are chosen to describe the galaxy
profiles. Elliptical galaxies are assumed to have a bulge-like profile
of $e^{-{r^{1/4}}}$, while disk galaxies have a disk-like profile of
$e^{-r}$.  Here, $r$ is the radial distance from the galaxy
center. Each profile is characterized by a major axis half--light
radius and axis ratio. Stellar (point-like) objects are examined
through the same procedure as the galaxy images, except a Gaussian
$e^{-r^{2}}$ profile is adopted.

For most of the galaxies, a 64--pixel square region around the center of
galaxy is examined. For larger galaxies, a 128--pixel square region is
chosen. The mask of detected objects (see above) is used to determine
a contour around each object which is $1 \sigma$ above the estimated
local sky. The total integrated signal-to-noise ratio of these pixels
is a good measure of the information content of the image. The
completeness limit of the object finding algorithm is $SNRIL \sim
1.5$, where $SNRIL$ is the decimal logarithm of the integrated
signal-to-noise ratio. However, only those galaxies with $SNRIL > 2$
have enough signal to be reliably fit to the full two-component model
discussed above (see ROG).
 
A full two--component model is fit to all objects with $SNRIL > 2$;
however, extensive testing has shown that galaxies with $SNRIL < 2$ do
not have sufficient signal to be fit to a model with such a large
number of parameters. Therefore, these images are examined only as a
pure disk, a pure bulge, or a stellar object. As a first pass, each
object is fit to a disk-like model if $SNRIL \le 2$ or a 10--parameter
disk $+$ bulge model if $SNRIL > 2$. If the half-light radius is less
than one pixel, a stellar profile is also fitted. The star--galaxy
classification is based on both the likelihood ratio for the best
galaxy model and the half-light radius. In addition, the following
checks are also made. Firstly, for those images originally fit by a
two--component disk $+$ bulge model, a single--component model with
fewer parameters is examined to determine whether the two--component
model is a significantly better fit. Secondly, for images with a
half-light radius less than two pixels, a single-component fit is
selected, as these galaxy images are too undersampled to give
realistic two--component models.
 
A maximum likelihood parameter estimation procedure is used to
determine the best model and parameter values. For each set of model
parameters, a model image of the object is created and compared with
the actual object image. The likelihood function is defined as the
product of the probabilities for each model pixel value with respect
to the observed pixel value and its error distribution. This function
is minimized by a routine described in Ratnatunga \& Casertano
(1991). Finally, a best-fit model and its parameters are determined
for each object with the following classifications : bulge $+$ disk,
disk, bulge, galaxy (if the classification as disk or bulge is not
significant), stellar, or object (if there is no preference between
star and galaxy). For the fine details on the parameter fittings and
the maximum likelihood estimator, see ROG and the references listed
above.
 
All of those objects classified as non--stellar down to the
completeness limit of $SNRIL > 1.5$ are presented; however, only those
galaxies with a reliable signal-to-noise ($SNRIL > 2$) are used in the
morphological analyses in Paper III. The magnitude to which this
signal-to-noise corresponds depends on the duration and passband of
the particular observation (see Table 3), e.g.\ a $SNRIL = 1.5$
corresponds to ${\rm F702W} \simeq 26.3$ and ${\rm F814W} \simeq 26.0$
for a stellar object in the CL0023+0423 and CL1604+4304 fields,
respectively. The catalogs and the first results from the automated
galaxy classifications of these two cluster fields are presented in
Paper III.
 
\subsubsection{Visual Classification of Galaxies in WFPC2 Images}
 
In addition to the automated classifications, we have also performed a
visual classification of the brightest galaxies detected in the MDS
automated object detection procedure described in Sect.\ 4.5.1.  For
each individual HST observation, we determine the total magnitude
limit down to which we can accurately classify galaxies by eye. We
define the total magnitude of each galaxy as the analytic total
magnitude of the best-fit galaxy model determined in the automated
classification procedure (see Sect.\ 4.5.2). The magnitude limit that
we choose depends on the specifics of the individual cluster
observation (e.g.\ ${\rm F702W} = 24.7$ for CL0023+0423 and ${\rm
F814W} = 24.3$ for CL1604+4304; see Paper III). These limits
correspond to $\sim 200$ galaxies per field.
 
We base our visual classifications on the Revised Hubble scheme, that
is, the standard galaxy classification systems used in the Hubble
Atlas (Sandage 1961) and the Carnegie Atlas of Galaxies (Sandage \&
Bedke 1994). The assignment of morphological types on the Revised
Hubble Type system is made by a visual inspection of images displayed
using IRAF and SAOIMAGE. A $60^{''} \times 60^{''}$ pixel grey scale
image centered on each galaxy brighter than our chosen magnitude limit
is examined.  Each image is stretched at 0--100DN and 0--400DN, using
a logarithmically scaled display. This display provides the wide
dynamic range necessary to detect both low surface brightness features
in the outer regions of the galaxy, as well as the high surface
brightness structure in the galaxy core. Furthermore, it most closely
mimics the response of photographic plates, which are the source of
the historical galaxy classification scheme. In addition to our
display techniques, we have used the CCD survey by Frei \etal (1996)
of 113 nearby galaxies as a reference sample in order to compare
directly our CCD images of high-redshift galaxies to those galaxies in
the local universe. These observations were made in either $B_{j}$ and
$R$ or $g$, $r$, and $i$. All of the Frei \etal galaxies have
classifications based on photographic plates (see the Carnegie Atlas
of Galaxies).
 
For the specifics of the galaxy classifications, we have adopted the
classification procedure first employed by Dressler \etal (1994) for
an HST study of galaxies in the field of the cluster CL0939+4713 at $z
= 0.41$, and adopted by Smail \etal (1997) for a comprehensive
catalog of morphological types in 10 intermediate redshift
clusters. This galaxy classification includes four components : (1)
Revised Hubble type, (2) disturbance index -- the perceived asymmetry
of the galaxy image, (3) dynamical state -- the interpretation of the
cause of any observed disturbance and (4) comments. The disturbance
index ($D$) ranges from 0 to 4 with the following definitions : 0 --
little or no asymmetry, 1 or 2 -- moderate or strong asymmetry, and 3
or 4 -- moderate or strong distortion.  This classification was
intended to be objective in that it is independent of the possible
reason of the disturbance. The ``dynamical state,'' however, was
intended to be a subjective and interpretive judgment of the cause of
the disturbance and should, therefore, be viewed only as an educated
guess. The classes assigned to this parameter are : I -- tidal
interaction with a neighbor, M -- tidal interaction suggesting a
merger, T -- tidal feature without obvious cause, and C --
chaotic. For more details on these four components, see Smail \etal
(1997) and Paper III.  All of the galaxies in our HST images are
classified by L.\ Lubin and M.\ Postman; in addition, A.\ Sandage
kindly provides expert classifications of the galaxy samples. In
particular, he reviews the rather tricky separation between elliptical
(E) and S0 galaxies. The independent classifications are merged by L.\
Lubin.

In Paper III, we present the results of our morphological study of the
clusters CL0023+0423 and CL1604+4304. This paper includes full
catalogs of the automated and visual galaxy classifications in these
fields, a comparison between the two classification methods, and a
study of the cluster galaxy populations.

\section{Summary}

We have described the data acquisition and reduction procedures of our
photometric and spectroscopic campaign to study nine candidate
clusters of galaxies at redshifts of $z > 0.6$. The observational
program consists of four main parts :

\newcounter{discnt}
\begin{list} {\arabic{discnt}.}
{\usecounter{discnt} \setlength{\leftmargin 0.2in}{\itemsep
0in}{\topsep 0in}{\parskip 0in}}

\item Spectra for approximately 80\% of all galaxies down to a
Johnson--Cousins $R$ magnitude of $\sim 23.5$ within a fixed area of
${2}^{'}.2 \times {7}^{'}.6$ of each cluster field using the
Low--Resolution Imaging Spectrograph (LRIS) at the Keck 10m telescope.
We have obtained spectra covering the range 4400 \AA\ to 9500 \AA\ for
$\sim 130 - 150$ galaxies per cluster field. Redshifts have been
determined for 892 objects.

\item Deep $BVRI$ imaging with Keck of all galaxies in the full LRIS
field of $6^{'} \times 8^{'}$ centered on each cluster.  The $5\sigma$ detection
limits are $B = 25.1$, $V = 24.1$, $R = 23.5$, and $I = 21.7$ in our
standard 3 arcsecond radius aperture.  The photometric data are
converted to absolute fluxes in order to obtain absolute spectral
energy distributions.

\item High angular resolution imagery with HST in order to provide
morphological information on the galaxies in the WFPC2 field-of-view
(${160}^{''} \times {160}^{''}$) centered on each cluster. Each of the
cluster candidates has been or will be observed by HST in Cycles 5 and
6.

\item High precision $K$ band photometry across the WFPC2
field--of--view with the KPNO 4m telescope for each cluster.  The
infrared survey reaches a limiting magnitude of $K^{'} = 20$ in
the standard aperture.
\end{list}

We have presented the redshift
histograms for the nine candidate clusters of galaxies in
this survey. We find that six of the nine
candidate clusters of galaxies are real density enhancements.  They
include CL0023+0423, CL0943+4804, CL1324+3011, CL1325+3009,
CL1604+4304, and CL1604+4321. 
The remaining three candidates are of a more dubious nature.
Their redshift distributions reveal no clear density enhancement but
rather an apparent superposition of small groups of galaxies along the
line-of-sight. This false positive rate is consistent with the
estimate of $\sim 30$\% provided in Postman \etal (1996). At lower redshifts
($z \simless 0.5$) the spurious rate is about 20\% or less. This
is based on spectroscopic follow-up of PDCS candidates 
being done by Holden \& Nichol (1998) at the KPNO 4m telescope. 
We conclude that optical detection of clusters remains a successful
and important method for identifying such systems out to $z \sim 1$
and, further, will provide an important complement to cluster searches
at other wavelengths.

Results on the star-formation history, dynamics, and morphological
properties of  CL0023+0423 ($z = 0.84$) and CL1604+4304 ($z = 0.90$)
are presented in Papers II and III.

\vskip 1.0cm
We thank Don Schneider and the anonomous referee for their 
invaluable comments on this manuscript.
The W.M. Keck Observatory is operated as a scientific partnership
between the California Institute of Technology, the University of
California, and the National Aeronautics and Space Administration.  It
was made possible by the generous financial support of the W. M. Keck
Foundation. LML graciously acknowledges support from a Carnegie
Fellowship. Support for this work was also provided, in part, by NASA
through grant number GO-06000.01-94A from the Space Telescope Science
Institute, which is operated by the Association of Universities for
Research in Astronomy, Inc., under NASA contract NAS5-26555.

\newpage
\pagestyle{empty}
\renewcommand{\arraystretch}{1.0}

\vfill
\begin{table}
\begin{center}
Table 1 : The Cluster Sample
\end{center}
\begin{center}
\scriptsize
\begin{tabular}[h]{clccl}
\hline
\hline
Cluster & Alternate ID & R.A. (2000)	& Dec. (2000)	& Redshifts   \\
\hline
CL0023+0423 &	00 21 18 +04 06 30	& 00 23 52.2 	& +04 23 07.4	& 0.63(6), 0.72(5), 0.84(25), 0.92(4), 1.10(4) \\
CL0231+0048 & 	02 29 09 +00 35 30	& 02 31 43.2	& +00 48 44.0	& 0.28(12), 0.58(9), 0.62(4), 0.75(5), 0.80(8), 1.00(9) \\
CL0943+4804 & 	09 40 25 +48 19 00	& 09 43 41.0	& +48 04 45.5	& 0.41(6), 0.47(13), 0.70(10), 0.94(4) \\
CL1324+3011 &	13 22 27 +30 27 18	& 13 24 50.4	& +30 11 25.8	& 0.59(4), 0.66(4), 0.76(15) \\
CL1325+3009 & 	PDCS S14		& 13 25 18.7	& +30 09 57.0	& 0.40(11), 0.62(7), 0.67(7), 0.72(12), 0.92(7), 1.07(11), 1.20(3) \\
CL1604+4304 & 	16 02 44 +43 12 54	& 16 04 19.5 	& +43 04 33.9	& 0.50(8), 0.83(7), 0.90(22) \\ 
CL1604+4321 & 	16 02 44 +43 29 24 	& 16 04 31.6	& +43 21 35.4	& 0.24(6), 0.46(5), 0.62(6), 0.70(8), 0.93(39)\\
CL1607+4109 & 	PDCS S23		& 16 07 30.9 	& +41 09 32.0	& 0.56(6), 0.60(6), 0.77(5), 0.92(3) \\
CL2157+0347 & 	21 55 19 +03 34 12	& 21 57 52.2	& +03 47 00.0   & 0.45(8), 0.64(5), 0.82(5), 0.85(4), 1.10(3) \\
\hline
\end{tabular}
\end{center}
\end{table}
\vskip 4.0in

\vfill
\begin{table}
\begin{center}
Table 2 : The Broad Band Photometric System
\end{center}
\begin{center}
\begin{tabular}[h]{cccccc}
\hline
\hline
	&	Mean		& Total			&\multicolumn{3}{c}{Galaxy at $z = 0.90$}	\\
Filter	&	Wavelength	& Exposure Time	& Age = 0.7 Gyr & Age = 1.4 Gyr & Age = 3.0 Gyr \\
	&	(\AA)		& (sec)			&\multicolumn{3}{c}{(${\rm AB} \pm \sigma_{\rm AB}$)}	\\
\hline
B	& 4416	& 3600 	& $22.30 \pm 0.02$	& $23.00 \pm 0.03$	& $25.10 \pm 0.23$ \\ 
V 	& 5458	& 2000  & $22.14 \pm 0.03$ 	& $22.60 \pm 0.05$ 	& $23.57 \pm 0.12$ \\
R 	& 6384	& 1200	& $21.87 \pm 0.04$ 	& $22.12 \pm 0.05$ 	& $22.51 \pm 0.07$ \\
I 	& 8424	& 900  	& $21.00 \pm 0.08$	& $21.00 \pm 0.08$	& $21.00 \pm 0.08$ \\
\hline
\end{tabular}
\end{center}
\end{table}

\newpage
\vfill
\begin{table}
\begin{center}
Table 3 : The HST Observations
\end{center}
\begin{center}
\begin{tabular}[h]{ccccl}
\hline
\hline
Cluster & Filter        &       Duration (ksec) & Cycle & Status        \\
\hline
CL0023+0423 &  F702W   & 17.9    & 5 & Observed Nov 1995 (GO-6000) \\
CL0231+0048 &  F606W   & 15.3    & 5 & Observed Nov 1995 (GO-6000) \\
CL0943+4804 &  F702W   & 18.8    & 6 & To be observed in 1998 (GO-6581) \\
CL1324+3011 &  F606W   & 16.0    & 4 & Archival data (PI Westphal; GO-5234) \\
            &  F814W   & 32.0    & 4 & Archival data (PI Westphal; GO-5234) \\
CL1325+3009 &  F814W   & 18.8    & 6 & To be observed in 1999 (GO-6581) \\
CL1604+4304 &  F814W   & 64.0    & 4 & Archival data (PI Westphal; GO-5234) \\
CL1604+4321 &  F702W   & 18.8    & 6 & Observed Dec 1997 (GO-6581) \\
CL1607+4109 &  F702W   & 18.6    & 5 & Observed Jul 1997 (GO-6000) \\
CL2157+0347 &  F702W   & 18.8    & 6 & Observed Jul 1997 (GO-6581) \\
\hline
\end{tabular}
\end{center}
\end{table}

\vfill
\eject
 
\end{document}